\documentclass[10pt,american,preprint, aps, nofootinbib]{revtex4-1}
\usepackage[T1]{fontenc}
\usepackage[latin9]{inputenc}
\usepackage{colortbl}
\usepackage{babel}
\usepackage{mathrsfs}
\usepackage{amsmath}
\usepackage{amssymb}
\usepackage[pdfusetitle,
 bookmarks=true,bookmarksnumbered=false,bookmarksopen=false,
 breaklinks=false,pdfborder={0 0 1},backref=false,colorlinks=false]
 {hyperref}

\makeatletter

\providecommand{\tabularnewline}{\\}

\usepackage{comment}
\usepackage{etoolbox}
\usepackage{breqn}

\makeatletter
\let\cat@comma@active\@empty
\makeatother

\newcommand{\breqnoverloadothers}
{%
    \renewenvironment{equation}{\ignorespaces\begin{dmath}}{\end{dmath}\ignorespacesafterend}%
    \renewenvironment{equation*}{\ignorespaces\begin{dmath*}}{\end{dmath*}\ignorespacesafterend}%
    \renewenvironment{multline}{\ignorespaces\begin{dmath}}{\end{dmath}\ignorespacesafterend}%
    \renewenvironment{multline*}{\ignorespaces\begin{dmath*}}{\end{dmath*}\ignorespacesafterend}%

}

\newcommand\breqnundefineothers
{%
    \renewenvironment{equation}{}{}%
    \renewenvironment{equation*}{}{}%
    \renewenvironment{multline*}{}{}%

}

\AtBeginEnvironment{dmath}{\breqnundefineothers}
\AtBeginEnvironment{dmath*}{\breqnundefineothers}

\AtBeginEnvironment{dgroup}{\def\breqnundefineothers{}\breqnoverloadothers}
\AtBeginEnvironment{dgroup*}{\def\breqnundefineothers{}\breqnoverloadothers}

\newcommand\brwrap[3]{%
  \setbox0=\hbox{$#2$}
  \left#1\vbox to \the\ht0{\hbox to 0pt{}}\right.\kern-.2em
  \begingroup #2\endgroup\kern-.15em
  \left.\vbox to \the\ht0{\hbox to 0pt{}}\right#3
}

\makeatother

\begin{document}
\title{Conformally Invariant Corrections to the Anomaly-Induced Effective
Action and Black Hole Evaporation in Four Dimensions}
\author{Bing-Nan Liu}
\author{David A. Lowe}
\affiliation{Department of Physics, Brown University, Providence, RI 02912, USA}
\begin{abstract}
When matter fields are integrated out in a large $N$ approximation,
the conformal anomaly induces an effective action up to conformally
invariant correction terms. In the present work, we consider the implications
on black hole evaporation of such a term involving a conformal invariant
found by Fefferman and Graham. Working in an approximation where the
spacetime is static, we compute the induced stress tensor around a
Schwarzschild black hole. The boundary conditions select an Unruh-like
asymptotic sector, but not a unique stress tensor. A new one-parameter
family of quantum hair emerges which changes the stress tensor in
the near-horizon region. This suggests a new semiclassical mechanism
by which information about black hole formation could be encoded outside
the horizon.
\end{abstract}
\maketitle

\section{Introduction}

The semiclassical description of black hole evaporation in four spacetime
dimensions remains an important test of effective field theory in
gravity. Hawking radiation is well understood as quantum field theory
on a fixed black hole background \citep{HAWKING_1974,Hawking1975},
but incorporating the corresponding back-reaction on the geometry
in a controlled and covariant approximation is considerably more difficult.
In two-dimensional models, anomaly-induced actions lead to tractable
local formulations in which the formation and evaporation of black
holes can be followed explicitly \citep{Callan:1992rs,Russo:1992ax}.
In four dimensions, the analogous problem is more intricate, but the
trace anomaly again provides a natural organizing principle for the
leading large-$N$ quantum effects of conformally coupled matter.

A useful starting point is Riegert\textquoteright s non-local effective
action \citep{RIEGERT198456}, which reproduces the four-dimensional
trace anomaly in a curved background. In earlier work, and in particular
in the generalized effective field theory treatment of four-dimensional
black hole evaporation, this action was augmented by the simplest
conformally invariant non-local correction involving the square of
the Weyl tensor \citep{Balbinot:1999ri,Balbinot:aa,Mazur2001,Mottola:2006ew,Anderson:2007aa,Liu:2025xfu},
see also \citep{Mottola:2016aa,Mottola:2025fhl} for related work.
After localization by auxiliary scalar fields, the resulting model
permitted an analytic computation of the induced stress tensor in
a Schwarzschild background \citep{Lowe:2025bxl}. With physically
motivated boundary conditions---regularity on the future horizon
and vanishing incoming flux at past null infinity---the static solution
was uniquely selected and had the expected qualitative properties
of the Unruh state. However, the model also exhibited two important
limitations. First, the Riegert contribution by itself gives the wrong
sign for the outgoing energy flux for ordinary conformally coupled
matter of spin $\leq1$. The Weyl-squared conformally invariant correction
introduced in the generalized model restored a positive luminosity.
Second, within that truncation, the physical boundary conditions removed
all integration constants in the stress tensor, leaving no analog
of quantum hair.

The purpose of the present paper is to investigate the next natural
step in this effective field theory expansion. The complete one-loop
effective action is expected to contain, in addition to the anomaly-determined
Riegert term, an infinite sequence of conformally invariant non-local
terms. These terms do not modify the trace anomaly, but they can affect
the semiclassical stress tensor and therefore the physics of black
hole evaporation. Here we include a new conformally invariant contribution
built from a weight-six invariant found by Fefferman and Graham \citep{fefferman}.
This term provides a controlled extension of the previous four-dimensional
effective field theory while preserving covariance and analytic tractability
in the static Schwarzschild problem.

We localize the extended non-local action by introducing two additional
auxiliary scalar fields, in addition to the two scalars appearing
in the earlier generalized model \citep{Liu:2025xfu}. The resulting
four-scalar formulation yields higher-derivative but local equations
of motion, which simplify substantially on the Schwarzschild background.
We solve these equations for static, spherically symmetric configurations
and compute the full anomaly-induced stress tensor. As in the earlier
work, we impose boundary conditions appropriate to an Unruh-like state:
absence of incoming radiation from past null infinity, finite outgoing
luminosity at future null infinity, and regularity for freely falling
observers crossing the future horizon.

The main result is that the Fefferman--Graham conformal correction
leaves the asymptotic Hawking flux under analytic control while introducing
a qualitatively new feature. The physical boundary conditions no longer
fix all parameters in the induced stress tensor. Instead, a one-parameter
family of regular solutions remains, with the free parameter affecting
the stress tensor near the black hole while falling off rapidly at
infinity. This parameter is not associated with the classical black
hole mass and does not change the imposed asymptotic flux conditions.
We therefore interpret it as a form of quantum hair generated by conformally
invariant terms in the anomaly-induced effective action. 

This quantum hair provides a new mechanism by which information about
black hole formation may be encoded in semiclassical observables outside
the horizon. In the static single-center setting considered here,
one may choose a particular value of the hair parameter as part of
the vacuum specification. In more general time-dependent or multi-centered
geometries, however, the same parameter may carry physical information
that is not visible in the anomaly alone. The present analysis therefore
suggests that conformally invariant sectors of the one-loop effective
action can play a more important role in four-dimensional evaporation
than is apparent from the trace anomaly by itself.

The paper is organized as follows. In Section II we construct the
anomaly-induced effective action supplemented by the Fefferman--Graham
conformal invariant and present its local scalar-tensor form. In Section
III we solve the auxiliary scalar equations on a Schwarzschild background.
In Section IV we impose the physical boundary conditions and derive
the resulting induced stress tensor and outgoing flux. In Section
V we analyze the remaining free parameter and interpret it as quantum
hair. We conclude in Section VI with a discussion of the implications
for semiclassical black hole evaporation and future time-dependent
studies.

\section{Anomaly Induced Action with Conformal Corrections}

Our starting point is the following general expression for the leading
order trace anomaly of the stress tensor in curved 3+1-dimensional
spacetime, for classically conformally coupled fields,
\begin{equation}
g^{ab}\left\langle T_{ab}\right\rangle =\frac{1}{16\pi^{2}}\left(a'C^{2}+b'E-c'\nabla^{2}R+d'F^{2}\right)\,,\label{eq:traceanom}
\end{equation}
where $a',b',c',d'$ are coefficients that depend on the matter content
of the theory\footnote{We use Misner-Thorne and Wheeler's $(+++)$ conventions \citep{misner}.}
and
\begin{align}
C^{2} & =R^{abcd}R_{abcd}-2R^{ab}R_{ab}+\frac{1}{3}R^{2},\\
E & =R^{abcd}R_{abcd}-4R^{ab}R_{ab}+R^{2},
\end{align}
are the square of the Weyl tensor and the Euler density, respectively.
The constant $d'$ is proportional to the beta function of the gauge
theory. In the present work we do not include gauge fields so set
$d'=0$. The $c'$ term may be removed by a local counterterm, however
we retain it in the present analysis (though as we will see, it does
not contribute to the induced stress tensor in a Schwarzschild background).

The anomaly coefficients, in terms of the number of matter species
with spins $\leq1$, are given by
\begin{align}
a' & =\frac{1}{120}\left(n_{s}+6n_{f}+12n_{V}\right),\nonumber \\
b' & =-\frac{1}{360}\left(n_{s}+11n_{f}+62n_{V}\right),\label{eq:abcd}\\
c' & =-\frac{1}{180}\left(n_{s}+6n_{f}+12n_{V}\right),\nonumber 
\end{align}
where $n_{s},n_{f}$ and $n_{V}$ are the number of scalars, Dirac
fermions and vectors respectively \citep{Duff:1993wm}. In order for
the semiclassical approximation to be under control the non-vanishing
coefficients are taken to be of order $N\gg1$ (while scaling $\hbar\sim1/N$)
so that the matter field contribution to the effective action is dominant
compared to that of the metric sector. 

The trace anomaly (with $d'=0$) is reproduced by a scalar-tensor
theory with two auxiliary scalar fields $\phi$ and $\chi$ with conformal
weight 0, as previously considered in \citep{Balbinot:1999ri,Balbinot:aa,Mazur2001,Mottola:2006ew,Anderson:2007aa}
and studied in the context of static black holes in \citep{Liu:2025xfu}.
In the following we include two additional scalar fields $\rho$ (weight
-2) and $\psi$ (weight 0) to include a new conformally invariant
term involving a weight 6 conformal invariant found by Fefferman and
Graham \citep{fefferman}, and a new weight 6 derivative term acting
on $\psi$ (a closely related discussion in the literature can be
found in \citep{Rachwal:2023oqp})

\begin{equation}
\mathcal{K}[\psi]=\tfrac{1}{3}R\nabla_{a}\psi\nabla^{a}\psi+\tfrac{1}{2}\nabla_{a}\nabla^{a}\psi\nabla_{b}\nabla^{b}\psi-R_{ab}\nabla^{a}\psi\nabla^{b}\psi\,,\label{eq:paneitz}
\end{equation}

\begin{dmath}
\begin{equation}
F_{6}=8R_{a}{}^{c}R^{ab}R_{bc}-\tfrac{28}{3}R_{ab}R^{ab}R+\tfrac{4}{3}R^{3}+8R^{ab}R^{cd}R_{acbd}-8R^{ab}R^{cd}W_{acbd}+\tfrac{2}{3}RW_{abcd}W^{abcd}-\nabla_{a}R\nabla^{a}R-12\nabla_{b}R_{ac}\nabla^{c}R^{ab}+12\nabla_{c}R_{ab}\nabla^{c}R^{ab}+16W_{acbd}\nabla^{d}\nabla^{c}R^{ab}+\nabla_{e}W_{abcd}\nabla^{e}W^{abcd}\,,\label{eq:fgterm}
\end{equation}
\end{dmath}

\begin{dmath}
\begin{equation}
\Delta_{6}[\psi]=-\tfrac{1}{24}R\nabla_{a}\psi\nabla^{a}R+\tfrac{1}{4}R^{bc}\nabla_{a}R_{bc}\nabla^{a}\psi+\tfrac{1}{24}R_{ab}\nabla^{a}R\nabla^{b}\psi-\tfrac{1}{2}W_{a}{}^{cde}W_{bcde}\nabla^{b}\nabla^{a}\psi-\tfrac{1}{4}R^{bc}\nabla^{a}\psi\nabla_{c}R_{ab}+\tfrac{1}{2}W^{bcde}\nabla^{a}\psi\nabla_{c}W_{abde}+\tfrac{1}{2}R^{bc}\nabla^{a}\psi\nabla_{d}W_{abc}{}^{d}+\tfrac{3}{2}W_{abcd}\nabla^{a}\psi\nabla^{d}R^{bc}\,,\label{eq:rachwal}
\end{equation}
\end{dmath}

\begin{dmath}
\begin{equation}
S=\int d^{4}x\,(-g)^{1/2}\left[\frac{1}{16\pi}R+\frac{1}{192\pi^{2}}(c'-\tfrac{2}{3}b')R^{2}-b'\mathcal{K}[\phi]-\mathcal{K}[\chi]+f_{2\psi}\mathcal{K}[\psi]-\frac{\phi}{8\pi}\left((a'+b')C^{2}+\frac{2b'}{3}\left(R^{2}-3R_{ab}R^{ab}-\nabla^{2}R\right)\right)+f_{4\chi}C^{2}\chi+f_{4\psi}C^{2}\psi+\rho\,F_{6}-\frac{1}{2}\rho\Delta_{6}[\psi]\right].\label{eq:action}
\end{equation}
\end{dmath} The terms involving the $\chi,\rho$ and $\psi$ fields
are conformally invariant and do not contribute to the trace anomaly.
The $\phi$-dependent Riegert term is fixed by the anomaly. The Weyl-squared
nonlocal correction dependent on $\chi$ is the lowest weight conformally
invariant term considered previously. The new terms involving $\psi$
and $\rho$ represent the next higher conformal-weight correction
at second order in the curvature.

Varying this action with respect to the metric leads to the Einstein
equations sourced by additional terms which we refer to as the induced
stress tensor. In the following we view the parameters $a',b',c'$
as fixed by the matter content of the theory, while the new parameters
$f_{2\psi}$, $f_{4\psi}$ and $f_{4\chi}$ are effective field theory
parameters. In principle, these might be fixed by truncating the full
one-loop action, though this is extremely complex, even for conformally
coupled scalars \citep{barvinsky2009covariantperturbationtheoryiv,Barvinsky:2023exr}.

The derivative operator \eqref{eq:paneitz} is related to the Paneitz
operator (by integration by parts) \citep{Paneitz_2008} and is the
unique conformal weight 4 operator that acts on weight 0 scalars.
The scalar invariant \eqref{eq:fgterm} was found in \citep{fefferman}
and has conformal weight 6. It is a conformally covariant improvement
of the term $|\nabla_{e}W_{abcd}|^{2}$ which is second order in the
Riemann tensor. It is unique up to the addition of the term third
order in the Weyl tensor $W^{\,\,cd}_{ab}W^{\,\,ef}_{cd}W^{\,\,ab}_{ef}$.
We choose not to include this higher order term, and focus on the
most general weight 6 terms second order in the Riemann tensor. The
derivative operator \eqref{eq:rachwal} is the unique weight 6 derivative
operator acting on weight 0 scalar fields, and is a generalization
of similar operators studied in \citep{Rachwal:2023oqp}. Interestingly,
it does not contain terms of the form $\square^{3}\psi$ so would
vanish in flat spacetime. The existence of apparently an infinite
sequence of such operators with conformal weights $\geq4$ is key
to the construction of this extension to the non-local Riegert action
proposed in the present work and in \citep{Liu:2025xfu}.

As a simple check on these expressions, one can take the trace of
the induced stress tensor derived from this action
\begin{equation}
g^{ab}\left\langle T_{ab}\right\rangle =\frac{b'}{2\pi}\left(\nabla^{2}\nabla^{2}\phi-\tfrac{2}{3}R\nabla^{2}\phi+2R^{ab}\nabla_{a}\nabla_{b}\phi+\tfrac{1}{3}\nabla^{a}R\:\nabla_{a}\phi\right)-\frac{1}{16\pi^{2}}\left(c'-\frac{2}{3}b'\right)\nabla^{2}R\,,\label{eq:stresstrace}
\end{equation}
and eliminate the scalar field using its equation of motion. Completing
this computation requires the use of some nontrivial Riemann tensor
identities which were carried out using the xAct MATHEMATICA package
\citep{Martin-Garcia:2007bqa,Martin-Garcia:2008ysv}. After some straightforward
algebra one then recovers the trace anomaly formula \eqref{eq:traceanom}
with $d'=0$. 

\section{Scalar Field Solutions}

The scalar equations of motion derived from the action \eqref{eq:action}
can be explicitly solved on the Schwarzschild black hole background,
\begin{equation}
ds^{2}=-\left(1-\frac{2M}{r}\right)dt^{2}+\frac{dr^{2}}{1-\frac{2M}{r}}+r^{2}d\Omega^{2}.\label{eq:schwarzschild}
\end{equation}
Since $R_{ab}=R=0$ in this background, the scalar equations reduce
to 
\begin{align}
\frac{(a'+b')R_{abcd}R^{abcd}}{8\pi}+b'\nabla_{b}\nabla^{b}\nabla_{a}\nabla^{a}\phi & =0\nonumber \\
f_{4\chi}{}R_{abcd}R^{abcd}-\nabla_{b}\nabla^{b}\nabla_{a}\nabla^{a}\chi & =0\nonumber \\
\tfrac{1}{16}R_{bcde}R^{bcde}\nabla_{a}\nabla^{a}\psi-\tfrac{1}{4}R^{bcde}\nabla^{a}\psi\nabla_{c}R_{abde}+\nabla_{e}R_{abcd}\nabla^{e}R^{abcd} & =0\nonumber \\
f_{4\psi}{}R_{abcd}R^{abcd}+\tfrac{1}{16}R_{bcde}R^{bcde}\nabla_{a}\nabla^{a}\rho+f_{2\psi}{}\nabla_{b}\nabla^{b}\nabla_{a}\nabla^{a}\psi-\tfrac{1}{4}R^{bcde}\nabla^{a}\rho\nabla_{c}R_{abde} & =0\,.
\end{align}
The general static spherically symmetric solution for each scalar
field, $\phi$ and $\chi$, can be written as a linear combination
of four independent solutions to the corresponding homogenous problem
plus special solutions, $\phi_{P}$ and $\chi_{P}$, that satisfy
the respective inhomogeneous equations,
\begin{align}
\phi & =a_{\phi1}\phi_{1}+a_{\phi2}\phi_{2}+a_{\phi3}\phi_{3}+a_{\phi4}+\phi_{P}\,,\nonumber \\
\chi & =a_{\chi1}\phi_{1}+a_{\chi2}\phi_{2}+a_{\chi3}\phi_{3}+a_{\chi4}+\chi_{P}\,,\label{eq:gensolution}
\end{align}
where
\begin{align}
\phi_{1} & =\log\left(1-\frac{2M}{r}\right)\,,\nonumber \\
\phi_{2} & =r^{2}+4Mr+8M^{2}\log r\,,\nonumber \\
\phi_{3} & =-\mathrm{Li}_{2}\left(\frac{2M}{r}\right)+\frac{r}{4M}+\frac{3}{2}\log r-\frac{1}{16M^{2}}\left(r^{2}+4Mr-8M^{2}\log r\right)\log\left(1-\frac{2M}{r}\right)\,,\nonumber \\
\phi_{P} & =\frac{(a'+b')}{8\pi b'}\Psi(r)\,,\nonumber \\
\chi_{P} & =-f_{4\chi}\,\Psi(r)\,,\nonumber \\
\Psi(r) & =\frac{2}{3M}r+2\log r+\frac{1}{12M^{2}}\left(r^{2}+4Mr+8M^{2}\log r\right)\log\left(1-\frac{2M}{r}\right)\,.\label{eq:solutions}
\end{align}

The two new scalars satisfy second order equations with static solutions\begin{dmath}
\begin{equation}
\psi=a_{\psi2}{}+\tfrac{1}{30}(a_{\psi1}{}-\frac{1}{4M^{5}})r(480M^{4}+120M^{3}r+40M^{2}r^{2}+15Mr^{3}+6r^{4})+40\log(r)+32a_{\psi1}{}M^{5}\log(-2M+r)\,,\label{eq:newsol}
\end{equation}
\end{dmath}and\begin{dmath}
\[
\rho=a_{\rho2}{}-\frac{f_{4\psi}{}r^{3}(40M^{2}+15Mr+6r^{2})}{45M^{3}}+\tfrac{1}{30}a_{\rho1}{}r(480M^{4}+120M^{3}r+40M^{2}r^{2}+15Mr^{3}+6r^{4})+\frac{f_{2\psi}{}r^{4}\bigl(r^{5}+M^{5}(48-4a_{\psi1}{}r^{5})\bigr)}{6M^{7}}+32a_{\rho1}{}M^{5}\log(-2M+r)\,.
\]
\end{dmath}The twelve constants $a_{Xi}$ (with $X=\phi,\chi,\rho,\psi)$
are to be fixed using physical boundary conditions. 

\section{The Induced Stress Tensor\label{sec:Stress-Tensor}}

Our goal is to evaluate the semiclassical stress tensor subject to
suitable boundary conditions applied to the asymptotic form of the
stress tensor as $r\to2M$ and $r\to\infty$ following the same logic
as \citep{Liu:2025xfu} and references therein.

We consider scalar fields of the form
\begin{align}
\phi(t,r) & =d_{\phi}\,t+\phi(r)\,,\label{eq:scalarsol}\\
\chi(t,r) & =d_{\chi}\,t+\chi(r)\,,\nonumber \\
\psi(t,r) & =d_{\psi}t+\psi(r)\,.
\end{align}
where $\phi(r),\chi(r)$ and $\psi(r)$ are static solutions of the
form \eqref{eq:gensolution} and \eqref{eq:newsol} and $d_{\phi},d_{\chi}$
and $d_{\psi}$ are additional free parameters to be fixed by boundary
conditions. This simple ansatz (considered originally in \citep{Mottola:2006ew})
gives a time-independent stress tensor. However the modification of
the stress tensor ends up depending on only 2 combinations of these
three parameters
\begin{align*}
\bar{d_{1}} & =d_{\psi}{}^{2}f_{2\psi}{}-{b'}d_{\phi}{}^{2}-d_{\chi}{}^{2}\,,\\
\bar{d}_{2} & =({a'}+{b'})d_{\phi}{}+2(3{b'}a_{\phi3}{}d_{\phi}{}+3a_{\chi3}{}d_{\chi}{}-4f_{4\chi}{}d_{\chi}{}-8d_{\psi}{}f_{4\psi}{}+12a_{\rho1}{}d_{\psi}{}M^{3})\pi\,.
\end{align*}
We then choose to solve for $\bar{d}_{1}$ and $\bar{d}_{2}$ to provide
a non-redundant solution for the stress tensor. We require the combinations
that enter the stress tensor to be real. Since the auxiliary fields
are not themselves observables, we do not separately impose a reality
condition on the constants $d_{\phi},d_{\chi}$ and $d_{\psi}$. For
non-vanishing $\bar{d}_{2}$ the solution breaks time-translation
invariance and gives rise to a non-vanishing $T_{rt}$ component. 

The full stress tensor is given by rather lengthy expressions that
we do not write out explicitly here (our MATHEMATICA expression contains
656 terms). The stress tensor is independent of $a_{\phi4}$, $a_{\chi4}$
and $a_{\psi2}$ so we begin by setting these parameters to zero.
In order that $T^{\theta}_{\theta}$ not grow at large $r$ we require
$a_{\psi1}=\frac{1}{4M^{5}}$. As we will see later, $a_{\rho2}$
will be undetermined by our physical constraints, and represents a
new quantum hair parameter. We will determine the impact of this in
the next section. For now, we focus on the remaining parameters and
set $a_{\rho2}=0$. Apart from the appearance of the new parameter
$a_{\rho1}$, the determination of the remaining constants follows
the same procedure as in the two-scalar model \citep{Liu:2025xfu}.

We are left with nine independent parameters to be determined. Some
of these parameters are fixed by requiring a freely falling observer
crossing the future horizon see a finite energy density. Some of the
same conditions come from requiring the $T_{\theta}{}^{\theta}$ component
of the stress tensor to be finite at the future horizon, which amounts
to a simpler calculation. Near the horizon,
\begin{equation}
T^{\,\,\theta}_{\theta}=\frac{A_{1}}{(r-2M)^{2}}+\frac{2A_{1}}{M(r-2M)}+A_{2}\log^{2}\left(\frac{r}{2M}-1\right)+A_{3}\log\left(\frac{r}{2M}-1\right)+\cdots\,,\label{eq:tthetatheta}
\end{equation}
while near infinity,
\begin{equation}
T^{\,\theta}_{\theta}=B_{1}+\frac{B_{2}}{r}+\frac{B_{3}}{r^{2}}+\frac{B_{4}}{r^{3}}+\cdots\,,\label{eq:tthth}
\end{equation}
where the $A_{i}$ and $B_{i}$ are quadratic functions of the $a_{Xi}$
and $\bar{d}_{1}$ and $\bar{d}_{2}$. As in the two-scalar model
there are 2 branches that produce the same stress tensor and are a
candidate for the Unruh-like vacuum state. These 2 branches appear
when $a_{\rho1}=0$. In addition there appear to be 6 other branches
arising from solving the set of simultaneous quadratic equations which
have $a_{\rho1}\neq0$. However these branches do not seem to be continuously
connected to the branches found in the two-scalar model. In appendix
\ref{sec:Quadratic-Constraint-Equations} we present the reduced constraint
equations for general $a_{\rho1}\neq0$ so their solution is determined
in terms of the remaining unknowns (which seems to be the most compact
way of presenting the solution, and may readily be solved numerically).
For now we focus on the branches with $a_{\rho1}=0$.

The conditions $B_{1}=B_{2}=B_{3}=0$ are solved by $a_{\phi2}=a_{\chi2}=0$.
With $a_{\phi2}\neq0$ and $a_{\chi2}\neq0$ there are additional
$\frac{\log r}{r^{3}}$ terms. Finiteness near the horizon requires
setting $A_{1}=A_{2}=A_{3}=0$. The $A_{2}$ coefficient involves
a sum of two squares and demanding $A_{2}=0$ on the space of real
parameters leads to two conditions,
\begin{equation}
a_{\phi3}=\frac{a'+b'}{6\pi b'}\,,\quad a_{\chi3}=-\frac{4f_{4\chi}}{3}\,.\label{eq:c3d3}
\end{equation}

Let us define the null geodesic vectors $n^{\mu}_{\pm}=\left(\frac{1}{1-\frac{2M}{r}},\pm1,0,0\right)$
in $(t,r,\theta,\phi)$ coordinates. The ingoing flux at past null
infinity $\mathscr{I}^{-}$ is 
\begin{equation}
\frac{1}{4}n^{\mu}_{+}T_{\mu\nu}n^{\nu}_{+}=\frac{C_{3}}{r^{2}}+\cdots\,.\label{eq:nullcontract}
\end{equation}
To find the analog of the Unruh state, with vanishing ingoing flux
we must set $C_{3}=0$. Our next requirement is $B_{4}=0$ which ensures
a faster than $r^{-3}$ falloff of the angular components of the stress
tensor, as is expected of the Unruh vacuum \citep{Christensen:1977jc}.
This fixes
\begin{align}
\bar{d}_{1} & =\frac{{a'}+{b'}-384(f_{4\psi}{}+40f_{2\psi}{})\pi^{2}}{32M^{2}\pi^{2}}\label{eq:dpvalue}\\
\bar{d}_{2} & =-\frac{({a'}+{b'})^{2}}{4{b'}M\pi}-\frac{16f_{4\chi}{}^{2}\pi}{M}\,.
\end{align}

With these constraints, we return to the near-horizon limit. Regularity
for a freely falling observer requires that 
\begin{align}
n^{\mu}_{-}T_{\mu\nu}n^{\nu}_{-} & =\frac{A_{4}}{(r-2M)^{3}}+\frac{A_{5}}{(r-2M)^{2}}+\frac{A_{6}}{r-2M}+\frac{A_{7}}{r-2M}\log\left(r-2M\right)+\\
 & A_{8}\log^{2}\left(r-2M\right)+A_{9}\log\left(r-2M\right)+\cdots\,,\nonumber 
\end{align}
be finite on the horizon, where again the coefficients $A_{4},\ldots,A_{9}$
are quadratic functions of $a_{Xi}$. 

Finally we solve the remaining near horizon conditions, $A_{1}=A_{5}=0,$
to fix $a_{\phi1}$ and $a_{\chi1}$. This leads to doubling, with
the solutions given by\begin{dmath}
\[
a^{\pm}_{\phi1}=\frac{64(a'+b')f_{4\psi}\pi\pm2Z^{1/2}/{b'}^{2}}{(a'+b')^{2}+64b'f_{4\chi}{}^{2}\pi^{2}}-\frac{(a'+b')\bigl(3+2\log(2M)\bigr)}{12b'\pi}
\]
\end{dmath}with\begin{dmath}
\[
Z={b'}^{3}f_{4\chi}{}^{2}\Bigl(-({a'}+{b'})^{3}({a'}+3{b'})-128{b'}({a'}+{b'})\bigl({a'}f_{4\chi}{}^{2}+2{b'}f_{4\chi}{}^{2}-10{a'}f_{4\psi}{}-10{b'}f_{4\psi}{}-248({a'}+{b'})f_{2\psi}{}\bigr)\pi^{2}-4096{b'}^{2}\bigl(f_{4\chi}{}^{4}+16f_{4\psi}{}^{2}-4f_{4\chi}{}^{2}(5f_{4\psi}{}+124f_{2\psi}{})\bigr)\pi^{4}\Bigr)
\]
\end{dmath}and \begin{dmath}
\[
a^{\pm}_{\chi1}=f_{4\chi}{}\left(\tfrac{4}{3}\log(2M)+2\right)-\frac{512{b'}f_{4\chi}{}f_{4\psi}{}\pi^{2}\mp\left(({a'}+{b'})Z^{1/2}\right)/\left(4{b'}^{2}f_{4\chi}{}\pi\right)}{({a'}+{b'})^{2}+64{b'}f_{4\chi}{}^{2}\pi^{2}}\,.
\]
\end{dmath} When the above solutions for $a_{Xi}$ are inserted,
all the coefficients $A_{i},B_{i}$ and $C_{i}$ vanish without imposing
any additional conditions.

These two different branches give the same quantum induced stress-tensor.
This indicates a degeneracy between the couplings of certain modes
of the scalars $\phi$ and $\chi$ for the special case of the Schwarzschild
background. In particular, they yield a prediction for the outgoing
null flux at future null infinity $\mathscr{I}^{+}$ , 
\begin{equation}
\frac{1}{4}n^{\mu}_{-}T_{\mu\nu}n^{\nu}_{-}=\frac{2f^{2}_{4\chi}}{M^{2}r^{2}}+\frac{\left(a'+b'\right)^{2}}{32\pi^{2}M^{2}b'\,r^{2}}+\cdots\,,\label{eq:outflux}
\end{equation}
corresponding to an object with finite outgoing luminosity. Recall
the parameters $a'$,$b'$ are known and depend on the matter content
of the theory while $f_{4\chi}$ is an effective field theory parameter
which we treat as a free parameter, but in principle is computable
assuming a truncation of the complete effective action of \citep{Barvinsky:2023exr}. 

The final result respects the asymptotic conditions described in table
1 of \citep{Christensen:1977jc}, assuming conformally coupled scalar
matter. 
\begin{table}

\begin{tabular}{|c|c|c|c|c|}
\hline 
\multicolumn{3}{|c|}{$r\to\infty$} & \multicolumn{2}{c|}{$r\to2M$}\tabularnewline
\hline 
\hline 
 & $T^{r}_{\,r}$ & $T^{\theta}_{\,\theta}$ & $T^{r}_{\,r}$ & $T^{\theta}_{\,\theta}$\tabularnewline
\hline 
CF & $\frac{1}{r^{2}}$ & $\frac{1}{r^{4}}$ & $-(1-\frac{2M}{r})^{-1}$ & $-1$\tabularnewline
\hline 
LL & $\frac{1}{r^{2}}$ & $\frac{1}{r^{4}}-\frac{\log r}{r^{4}}$ & $-(1-\frac{2M}{r})^{-1}$ & $-1$\tabularnewline
\hline 
\end{tabular}\caption{Asymptotic behavior of the stress tensor expectation values, with
numerical factors omitted. The bottom row indicates the results of
the present paper, while the row above shows the expectations of Christensen
and Fulling \citep{Christensen:1977jc}.}

\end{table}
 In particular, with $f_{4\chi}$ suitably chosen, the outgoing flux
at infinity is positive, in line with Hawking's prediction. Likewise,
one has vanishing ingoing flux at past null infinity. However, the
$r^{-4}\log r$ behavior at large $r$ indicates the four-scalar model
does not give a perfect match to the expected result, as already noted
in studies of the two-scalar model \citep{Liu:2025xfu}.

\section{Quantum Hair\label{sec:Quantum-Hair}}

We interpret the scalars $\phi,\rho,\chi$ and $\psi$ as auxiliary
fields, so the only observables associated with them are derived from
the metric itself. A novel feature of the present construction is
that physical asymptotic conditions on the metric and the induced
stress tensor at infinity and on the future horizon do not fix the
parameter $a_{\rho2}$, other than requiring finiteness. Thus this
parameter can be interpreted as a new form of quantum hair. Note in
earlier discussions of the one-scalar and two-scalar models there
were no analogous parameters, and the asymptotic conditions completely
fixed the vacuum state. The observable hair is not the value of the
auxiliary scalar itself, but the induced stress tensor deformation
that remains after all asymptotic and horizon regularity conditions
have been imposed. The stress tensor contribution due to $a_{\rho2}$
takes the diagonal form
\begin{align*}
\delta T_{tt} & =-\frac{96M(2M-r)(91M^{2}-85Mr+20r^{2})}{r^{10}}a_{\rho2}\\
\delta T_{rr} & =\frac{96M(21M^{2}-20Mr+5r^{2})}{(2M-r)r^{8}}a_{\rho2}\\
\delta T_{\theta\theta} & =\frac{48M(112M^{2}-105Mr+25r^{2})}{r^{7}}a_{\rho2}\,,
\end{align*}
which falls off like a rapid power law near infinity, and is also
smooth on the future horizon
\[
n^{\mu}_{-}\delta T_{\mu\nu}n^{\nu}_{-}=\frac{480M(-7M+3r)}{r^{8}}a_{\rho2}\,.
\]
Thus the addition of the weight-six conformally invariant term in
the effective action provides an explicit semiclassical mechanism
for evading the usual no-hair theorems \citep{Heusler_1996}. For
a single black hole, one might argue $a_{\rho2}$ is simply fixed
by demanding vacuum boundary conditions on past null infinity $\mathcal{I}^{-}$.
However the parameter only influences the induced stress energy in
the vicinity of the black hole, so one could imagine a multi-centered
black hole solution where the analog of $a_{\rho2}$ approached different
constant values near the different horizons. Thus we argue that $a_{\rho2}$
represents a new type of quantum hair that can encode information
about the formation of the black hole. Moreover, within the present
static effective theory, the boundary conditions do not fix its magnitude,
and there is no apparent parametric suppression that would force it
to be negligible. This opens the possibility that conformally invariant
sectors of the one-loop effective action can leave observable imprints
in near-horizon black hole physics. We plan to study this phenomenon
further in the context of time-dependent solutions in the future. 

\section{Discussion}

In this paper we have extended the generalized effective field theory
description of four-dimensional black hole evaporation by adding a
new conformally invariant term to the anomaly-induced effective action.
The term is built from the Fefferman--Graham weight-six conformal
invariant and represents a further contribution beyond the anomaly-determined
Riegert action and the Weyl-squared correction studied previously
\citep{Liu:2025xfu}. Since conformally invariant terms do not affect
the trace anomaly, their coefficients are not fixed by the anomaly
coefficients. Nevertheless, they contribute nontrivially to the semiclassical
stress tensor and therefore to the physical evaporation problem.

After localizing the extended non-local action with auxiliary scalar
fields, we obtained a four-scalar model whose equations can be solved
analytically in a Schwarzschild background. Imposing the usual Unruh-type
boundary conditions---regularity on the future horizon, and vanishing
incoming flux at past null infinity---again selects a controlled
semiclassical state. The outgoing luminosity contains the anomaly-induced
contribution together with a positive contribution governed by the
coefficient of the conformally invariant Weyl-squared sector. Thus,
as in the preceding generalized effective field theory model, the
sign problem of the minimal Riegert action for ordinary spin $\leq1$
conformal matter can be avoided without invoking exotic matter species.

The new feature of the present construction is that the physical boundary
conditions do not uniquely fix the full induced stress tensor. A single
parameter remains, producing a finite, diagonal correction to the
stress tensor that is smooth on the future horizon and decays rapidly
at large radius. This parameter does not alter the imposed asymptotic
radiation conditions and is not determined by the black hole mass.
It therefore represents a genuine form of quantum hair in the semiclassical
effective theory. Its origin lies in the conformally invariant part
of the one-loop effective action, rather than in the trace anomaly
itself.

This result sharpens the lesson of the earlier work. The minimal anomaly-induced
action, and the first conformally invariant extension, led to a unique
Unruh-like stress tensor once the physical boundary conditions were
imposed. The inclusion of the Fefferman--Graham invariant shows that
this uniqueness is not generic in the full effective field theory.
Conformally invariant terms that are invisible to the trace anomaly
can leave finite, observable imprints near the horizon. In this sense,
the trace anomaly fixes only part of the semiclassical physics; the
remaining conformally invariant sector can carry additional state-dependent
information.

The quantum hair found here is especially suggestive for the black
hole information problem. In the static Schwarzschild calculation,
the hair parameter may simply be chosen as part of the specification
of the state. In a dynamical collapse geometry, however, its value
could be determined by the history of the formation process. In multi-centered
or time-dependent situations, different near-horizon regions may support
different values of the analogous parameter while preserving the same
asymptotic flux conditions. This raises the possibility that information
about the initial state is encoded in semiclassical stress-energy
outside the horizon through conformally invariant terms in the effective
action. Since this stress energy is generated by quantum effects and
is localized in a region that is potentially observable, the corresponding
hair parameter may provide a channel through which near-horizon quantum
gravity effects leave observable imprints on astrophysical black holes. 

Several limitations of the present analysis should be emphasized.
We have restricted attention to a static Schwarzschild background
and have not included the back-reaction of the induced stress tensor
on the geometry. The resulting Unruh-like state has a constant outgoing
flux and therefore cannot by itself describe a finite-mass evaporating
black hole for all time. Thus the present calculation should be viewed
as a diagnostic of the local and asymptotic structure of the semiclassical
stress tensor, not as a complete evaporation solution. In addition,
the model studied here is only a finite truncation of the full conformally
invariant part of the one-loop effective action. 

Depending on the ranges of the effective field-theory parameters $f_{2\psi}\ensuremath{,}f_{4\chi}$
and $f_{4\psi}$, different branches of the Unruh state analogue may
give non-physical complex solutions to $a_{Xi}$. Even for real values,
these additional branches may generally be unstable (aside from the
branch noted in the main body of the text). It is possible that the
stability of each branch changes across the parameter space, suggesting
the emergence of black hole phase transitions. 

The next step is to study the corresponding time-dependent semiclassical
equations with back-reaction included. In the one-scalar model, this
has already been accomplished in \citep{Lowe:2026kvi} and without
back-reaction in \citep{Lowe:2025tik}. Such solutions would determine
whether the quantum hair identified in the static analysis is dynamically
generated during collapse, whether it remains stable during evaporation,
and whether it affects the late-time outgoing radiation. More generally,
the results of this paper indicate that a complete effective field
theory treatment of four-dimensional black hole evaporation must include
not only the anomaly-determined Riegert action but also the conformally
invariant sector of the non-local effective action. That sector may
provide the additional semiclassical structure needed to understand
how information is stored and released in four-dimensional black hole
evaporation.
\begin{acknowledgments}
We thank J. Hudson and L. Thorlacius for helpful discussions.
\end{acknowledgments}

\appendix

\section{Constraint Equations\label{sec:Quadratic-Constraint-Equations}}

Solving the general constraint equations imposes the conditions
\begin{equation}
a_{\psi1}=\frac{1}{4M^{5}},\ a_{\psi2}=0\,.
\end{equation}
The interpretation of $a_{\rho2}$ is detailed in section \ref{sec:Quantum-Hair}.
The solutions then further break up into a set of solutions with $a_{\phi2}=0,\:a_{\chi2}=0$
and another special branch with $\ a_{\phi2}\neq0,\:a_{\chi2}\neq0\,,$
but with $\bar{d}_{2}=0$. We present the reduced constraint equations
for these two main cases which is the most compact way of presenting
the solutions. The equations are arranged into the maximal set of
linear equations together with the remaining irreducible quadratic
constraints.

\subsection{Branch with $a_{\phi2}=0,\:a_{\chi2}=0$ }

There remain 7 independent constraints for the 7 remaining variables
$a_{\phi1},a_{\chi1},a_{\phi3},$ $a_{\chi3},\bar{d_{1}},\bar{d}_{2}$
and $a_{\rho1}$. As described above, there are 4 branches to these
solutions. Two have been given explicitly above, and are continuously
connected to the branches of the two-scalar setup \citep{Liu:2025xfu}.
The remaining 2 branches lead to simultaneous quadratic equations
that generate rather unwieldy expressions that we do not present in
this paper, and generate values for the Hawking radiation distinct
from \eqref{eq:outflux}. For completeness we present these 7 reduced
constraints, which we have arranged into 4 linear constraints and
3 irreducible quadratic constraints: \begin{dmath}
\[
3({a'}+{b'})a_{\phi3}{}+2\bar{d}_{2}{}M-24a_{\chi3}{}f_{4\chi}{}\pi=0\,,
\]
\end{dmath}\begin{dmath}
\[
55({a'}+{b'})^{2}-330{b'}({a'}+{b'})a_{\phi3}{}\pi+16{b'}\bigl(55f_{4\chi}{}(3a_{\chi3}{}+4f_{4\chi}{})-3618a_{\rho1}{}M^{3}\bigr)\pi^{2}=0\,,
\]
\end{dmath}\begin{dmath}
\[
-4({a'}+{b'})(7{a'}+10{b'})+24{b'}\bigl(12({a'}+{b'})a_{\phi1}{}+(7{a'}+10{b'})a_{\phi3}{}-12\bar{d}_{2}{}M\bigr)\pi-\tfrac{64}{5}{b'}\bigl(5f_{4\chi}{}(36a_{\chi1}{}+21a_{\chi3}{}+28f_{4\chi}{})+1440f_{4\psi}{}+3312a_{\rho1}{}M^{3}\bigr)\pi^{2}+24\bigl(({a'}+{b'})^{2}+6{b'}({a'}+{b'})a_{\phi3}{}\pi+16{b'}f_{4\chi}{}(-3a_{\chi3}{}+4f_{4\chi}{})\pi^{2}\bigr)\log(2M)=0\,,
\]
\end{dmath}\begin{dmath}
\[
-9{a'}+576\bigl(12f_{4\psi}{}+480f_{2\psi}{}+M^{2}(\bar{d}_{1}{}-18a_{\rho1}{}M)\bigr)\pi^{2}-9{b'}(1+6a_{\phi3}{}\pi)=0\,,
\]
\end{dmath}\begin{dmath}
\[
({a'}+{b'})^{2}+4{b'}({a'}+{b'})(4a_{\phi1}{}-3a_{\phi3}{})\pi+4{b'}\Bigl({b'}(4a_{\phi1}{}-3a_{\phi3}{})^{2}+(-4a_{\chi1}{}+3a_{\chi3}{}+4f_{4\chi}{})^{2}+64\bigl(-16f_{2\psi}{}+M^{2}(\bar{d}_{1}{}-48a_{\rho1}{}M)\bigr)\Bigr)\pi^{2}\biggr)+4\log(2M)\biggl(3\Bigl(({a'}+{b'})^{2}+8{b'}({a'}+{b'})a_{\phi1}{}\pi+4{b'}\bigl(3{b'}(4a_{\phi1}{}-3a_{\phi3}{})a_{\phi3}{}+(4a_{\chi1}{}-3a_{\chi3}{}-4f_{4\chi}{})(3a_{\chi3}{}-4f_{4\chi}{})\bigr)\pi^{2}\Bigr)+\Bigl(({a'}+{b'})^{2}+12{b'}({a'}+{b'})a_{\phi3}{}\pi+4{b'}\bigl(9{b'}a_{\phi3}{}^{2}+(3a_{\chi3}{}-4f_{4\chi}{})^{2}\bigr)\pi^{2}\Bigr)\log(2M)=0\,,
\]
\end{dmath}\begin{dmath}
\[
-11({a'}+{b'})^{2}+12{b'}({a'}+{b'})(5a_{\phi1}{}+a_{\phi3}{})\pi+4{b'}\bigl(9{b'}a_{\phi3}{}(-10a_{\phi1}{}+9a_{\phi3}{})+(-30a_{\chi1}{}+27a_{\chi3}{}-44f_{4\chi}{})(3a_{\chi3}{}+4f_{4\chi}{})+1728a_{\rho1}{}M^{3}\bigr)\pi^{2}+5\bigl(({a'}+{b'})^{2}-4{b'}(9{b'}a_{\phi3}{}^{2}+9a_{\chi3}{}^{2}-16f_{4\chi}{}^{2})\pi^{2}\bigr)\log(2M)=0\,,
\]
\end{dmath}\begin{dmath}
\[
{a'}^{2}+2{a'}{b'}(1-6a_{\phi3}{}\pi)+{b'}\bigl(4(3a_{\chi3}{}+4f_{4\chi}{})^{2}\pi^{2}+{b'}(1-6a_{\phi3}{}\pi)^{2}\bigr)=0\,.
\]
\end{dmath}The outgoing flux on this branch is given by \eqref{eq:outflux}
together with the two additional solutions\begin{dmath}
\[
\frac{1}{4}n^{\mu}_{-}T_{\mu\nu}n^{\nu}_{-}=-\frac{1}{8{b'}M^{2}\pi^{2}\bigl((103460{a'}+75119{b'})^{2}+685054182400{b'}f_{4\chi}{}^{2}\pi^{2}\bigr)r^{2}}\left((1300{a'}-51161{b'})({a'}+{b'})^{2}(103460{a'}+75119{b'})+64{b'}\Bigl(413840{a'}^{2}\bigl(650f_{4\chi}{}^{2}+603(81f_{4\psi}{}+8308f_{2\psi}{})\bigr)+4{a'}{b'}\bigl(-1231616590f_{4\chi}{}^{2}+107683137(81f_{4\psi}{}+8308f_{2\psi}{})\bigr)+{b'}^{2}\bigl(-5998674641f_{4\chi}{}^{2}+181187028(81f_{4\psi}{}+8308f_{2\psi}{})\bigr)\Bigr)\pi^{2}+1695088640{b'}^{2}f_{4\chi}{}^{2}\bigl(325f_{4\chi}{}^{2}+603(81f_{4\psi}{}+8308f_{2\psi}{})\bigr)\pi^{4}\pm4081104(-{b'})^{3/2}\pi f_{4\chi}{}\left(5820{a'}^{2}+7151{a'}{b'}+{b'}\bigl(1331{b'}+128(2910f_{4\chi}{}^{2}+5427f_{4\psi}{}+556636f_{2\psi}{})\pi^{2}\bigr)\right)\right)\,.
\]
\end{dmath}As one can see from this expression, while analytic solutions
of these equations are relatively straightforward, they quickly lead
to very long formulas. 

\subsection{Branch with $\bar{d}_{2}=0$}

The remaining branch has $\bar{d}_{2}=0$ and therefore has no outgoing
Hawking radiation. There remain 8 independent constraints for the
8 remaining variables $a_{\phi1},a_{\chi1},a_{\phi2},a_{\chi2},a_{\phi3},$
$a_{\chi3},\bar{d_{1}}$ and $a_{\rho1}$. This is a regular zero-flux
state, analogous in its asymptotic flux properties to the Boulware
state \citep{Boulware:1974dm} , that is allowed by the truncated
effective action. The reduced constraint equations are 3 linear constraints
and 5 irreducible quadratic constraints which lead to 4 different
branches corresponding to different stress tensors
\begin{equation}
({a'}+{b'})a_{\phi3}{}-8a_{\chi3}{}f_{4\chi}{}\pi=0\,,
\end{equation}
\begin{dmath}
\[
55({a'}+{b'})^{2}-66{b'}({a'}+{b'})(5a_{\phi3}{}-24a_{\phi2}{}M^{2})\pi+16{b'}\bigl(55f_{4\chi}{}(3a_{\chi3}{}+4f_{4\chi}{})-792a_{\chi2}{}f_{4\chi}{}M^{2}-3618a_{\rho1}{}M^{3}\bigr)\pi^{2}=0\,,
\]
\end{dmath}\begin{dmath}
\[
-5({a'}+{b'})(7{a'}+10{b'})+6{b'}\bigl(60({a'}+{b'})a_{\phi1}{}+5(7{a'}+10{b'})a_{\phi3}{}-384({a'}+{b'})a_{\phi2}{}M^{2}\bigr)\pi-16{b'}\bigl(5f_{4\chi}{}(36a_{\chi1}{}+21a_{\chi3}{}+28f_{4\chi}{})+1440f_{4\psi}{}-1152a_{\chi2}{}f_{4\chi}{}M^{2}+3312a_{\rho1}{}M^{3}\bigr)\pi^{2}+30\bigl(({a'}+{b'})^{2}+6{b'}({a'}+{b'})a_{\phi3}{}\pi+16{b'}f_{4\chi}{}(-3a_{\chi3}{}+4f_{4\chi}{})\pi^{2}\bigr)\log(2M)=0\,,
\]
\end{dmath}\begin{dmath}
\[
({a'}+{b'})^{2}+4{b'}({a'}+{b'})(4a_{\phi1}{}-3a_{\phi3}{})\pi+4{b'}\Bigl({b'}(4a_{\phi1}{}-3a_{\phi3}{})^{2}+(-4a_{\chi1}{}+3a_{\chi3}{}+4f_{4\chi}{})^{2}+64\bigl(-16f_{2\psi}{}+M^{2}(\bar{d}_{1}{}-48a_{\rho1}{}M)\bigr)\Bigr)\pi^{2}\biggr)+4\log(2M)\biggl(3\Bigl(({a'}+{b'})^{2}+8{b'}({a'}+{b'})a_{\phi1}{}\pi+4{b'}\bigl(3{b'}(4a_{\phi1}{}-3a_{\phi3}{})a_{\phi3}{}+(4a_{\chi1}{}-3a_{\chi3}{}-4f_{4\chi}{})(3a_{\chi3}{}-4f_{4\chi}{})\bigr)\pi^{2}\Bigr)+\Bigl(({a'}+{b'})^{2}+12{b'}({a'}+{b'})a_{\phi3}{}\pi+4{b'}\bigl(9{b'}a_{\phi3}{}^{2}+(3a_{\chi3}{}-4f_{4\chi}{})^{2}\bigr)\pi^{2}\Bigr)\log(2M)=0\,,
\]
\end{dmath}\begin{dmath}
\[
-11({a'}+{b'})^{2}+12{b'}({a'}+{b'})(5a_{\phi1}{}+a_{\phi3}{}-24a_{\phi2}{}M^{2})\pi+4{b'}\bigl((-30a_{\chi1}{}+27a_{\chi3}{}-44f_{4\chi}{})(3a_{\chi3}{}+4f_{4\chi}{})+144M^{2}(3a_{\chi2}{}a_{\chi3}{}+4a_{\chi2}{}f_{4\chi}{}+12a_{\rho1}{}M)+9{b'}a_{\phi3}{}(-10a_{\phi1}{}+9a_{\phi3}{}+48a_{\phi2}{}M^{2})\bigr)\pi^{2}+5\bigl(({a'}+{b'})^{2}-4{b'}(9{b'}a_{\phi3}{}^{2}+9a_{\chi3}{}^{2}-16f_{4\chi}{}^{2})\pi^{2}\bigr)\log(2M)=0\,,
\]
\end{dmath}\begin{dmath}
\[
{a'}^{2}+2{a'}{b'}(1-6a_{\phi3}{}\pi)+{b'}\bigl(4(3a_{\chi3}{}+4f_{4\chi}{})^{2}\pi^{2}+{b'}(1-6a_{\phi3}{}\pi)^{2}\bigr)=0\,,
\]
\end{dmath}
\begin{equation}
{b'}a_{\phi2}{}^{2}+a_{\chi2}{}^{2}=0\,,
\end{equation}
\begin{dmath}
\[
{a'}a_{\phi2}{}+2a_{\chi2}{}(3a_{\chi3}{}-4f_{4\chi}{}+8a_{\chi2}{}M^{2})\pi+{b'}a_{\phi2}{}(1+6a_{\phi3}{}\pi+16a_{\phi2}{}M^{2}\pi)=0\,.
\]
\end{dmath}

\bibliographystyle{utcaps}
\bibliography{riegert}

\end{document}